\documentclass[11pt,twoside]{article}
\usepackage{asp2006}
\usepackage{graphicx}
\usepackage{lscape}
\def\edcomment#1{\iffalse\marginpar{\raggedright\sl#1\/}\else\relax\fi}
\reversemarginpar

\begin{document}
\title{Optically-passive spirals: The missing link in\\ gradual star formation suppression upon cluster infall}

\begin{quote}
Christian Wolf,$^1$
Alfonso Arag\'on-Salamanca,$^2$
Michael Balogh,$^{3}$
Marco Barden,$^4$
Eric F. Bell,$^5$
Meghan E. Gray,$^2$
Chien Y. Peng,$^{6}$
David Bacon,$^7$
Fabio D. Barazza,$^{8}$
Asmus B\"ohm,$^{9}$
John A.R. Caldwell,$^{10}$
Anna Gallazzi,$^5$
Boris H\"au\ss ler,$^2$
Catherine Heymans,$^{11}$
Knud Jahnke,$^{5}$
Shardha Jogee,$^{12}$
Eelco van Kampen,$^4$
Kyle Lane,$^2$
Daniel H. McIntosh,$^{13}$
Klaus Meisenheimer,$^{5}$
Casey Papovich,$^{14}$
Sebastian F. S\'anchez,$^{15}$
Andy Taylor,$^{11}$
Lutz Wisotzki,$^{9}$
Xianzhong Zheng$^{16}$\\
\footnotesize
{\itshape $^1$Department of Physics, University of Oxford, Oxford, OX1 3RH, UK}\\
{\itshape $^2$University of Nottingham, Nottingham, NG7 2RD, UK}\\
{\itshape $^{3}$University Of Waterloo, Waterloo, Ontario, N2L 3G1, Canada}\\
{\itshape $^4$University of Innsbruck, Technikerstr.~25/8, 6020 Innsbruck, Austria}\\
{\itshape $^5$MPI f\"{u}r Astronomie, K\"{o}nigstuhl 17, 69117, Heidelberg, Germany}\\
{\itshape $^6$NRC Herzberg Institute of Astrophysics, 5071, Victoria, V9E 2E7, Canada}\\
{\itshape $^7$ICG, University of Portsmouth, Hampshire Terrace, Portsmouth, PO1 2EG, UK}\\
{\itshape $^8$EPFL, Observatoire de Sauverny, CH-1290 Versoix, Switzerland}\\
{\itshape $^{9}$AIP, An der Sternwarte 16, 14482 Potsdam, Germany}\\
{\itshape $^{10}$University of Texas, McDonald Observatory, Fort Davis, TX 79734, USA}\\
{\itshape $^{11}$SUPA, IfA, University of Edinburgh, Blackford Hill, Edinburgh, EH9 3HJ, UK}\\
{\itshape $^{12}$University of Texas, 1 University Station, Austin, TX 78712-0259, USA}\\
{\itshape $^{13}$University of Missouri, 5110 Rockhill Rd, Kansas City, MO 64110, USA}\\
{\itshape $^{14}$Texas A\&M University, College Station, TX 77843, USA}\\
{\itshape $^{15}$CAHA (Calar Alto), C/Jesus Durban Remon 2-2, 04004 Almeria, Spain}\\
{\itshape $^{16}$Purple Mountain Observatory, Ch.~Acad.~of Sc., Nanjing 210008, PR China.}\\
\normalsize
\end{quote}

\begin{abstract}
Galaxies migrate from the blue cloud to the red sequence when their star formation is quenched. Here, we report on galaxies quenched by environmental effects and not by mergers or strong AGN as often invoked: They form stars at a reduced rate which is optically even less conspicuous, and manifest a transition population of blue spirals evolving into S0 galaxies. These 'optically passive' or 'red spirals' are found in large numbers in the STAGES project (and by Galaxy Zoo) in the infall region of clusters and groups. 
\end{abstract}

\vspace{-0.5cm}
\section{The question}

This meeting has already discussed the transformation of galaxies from blue disk-like star-forming galaxies to red, spheroidal and quiescent galaxies. We have discussed the role of AGN in triggering star formation as well as suppressing it, and the role of mergers in intiating this phenomenon. Also, it was suggested that counts of mergers or AGN may correlate with stellar mass growth on the red sequence. In this paper, I will report on a category of galaxy that appears to undergo star formation quenching, while definitely not undergoing major mergers and perhaps not an AGN period either. They enrich the red sequence without the conspicuous signposts of galaxy transformation. Instead, they highlight the role of environment in transforming galaxies.

We already know that at low redshift both star formation and nuclear activity decline towards regions of higher galaxy density, even though the galaxy mass dependence of the two processes is very different. It is reasonable to expect that at sufficiently high redshift these trends have been inverted, and \citet{Wolf_El07} already found evidence for this. The connection between the two has yet to be clarified, which is of course the main question of this conference series. However, star formation is expected to be regulated by a range of processes, of which some are internal to a galaxy such as winds and ionisation from stars, X-ray binaries, SNe and AGN; others originate from the external environment, imposed onto a galaxy by dynamical interaction with other galaxies or a cluster potential, or by mechanical interaction with hot intra-cluster/group gas.

An obvious question in the study of galaxy transformation concerns the intermediate population of galaxies showing the progress of evolution in an unfinished state and giving away clues to the physics of the events. As far as star-formation quenching is concerned, the identification of transition objects has not been straightforward. Their apparent absence argued for a fast process leaving it unlikely to catch an event in progress \citep{Wolf_Ba04}, while the properties of most quiescent galaxies favour slow quenching, e.g.~the rarity of strong H$\delta$ absorption. Morphological change is clearly slower, and ÔanemicÕ or Ôoptically passiveÕ spirals with red colour and smooth arms have been known for long \citep{Wolf_vdB76,Wolf_Po99}.

In the following, I present new results on red spirals from the STAGES project \citep{Wolf_G09}, which combines optical SEDs and redshifts from COMBO-17 with ACS imaging from the largest contiguous HST mosaic after COSMOS, and Spitzer $24\mu$m data. The data cover $30\arcmin\times 30\arcmin$ or $5\times 5$~Mpc$^2$ at the redshift $z=0.165$ of the target cluster complex Abell 901/2, which is comprised mainly of four cores but includes further filaments and infalling groups. The cluster sample has very low field contamination of $\sim 10$\% given their extremely low photo-z scatter of $0.005$ rms. While the STAGES project includes data from further wavelength domains such as X-ray and radio, I will only report on the optical and $24\mu$m data, and will thus not comment on AGN activity either.

\begin{figure*}
\centering
\includegraphics[clip,angle=270,width=\hsize]{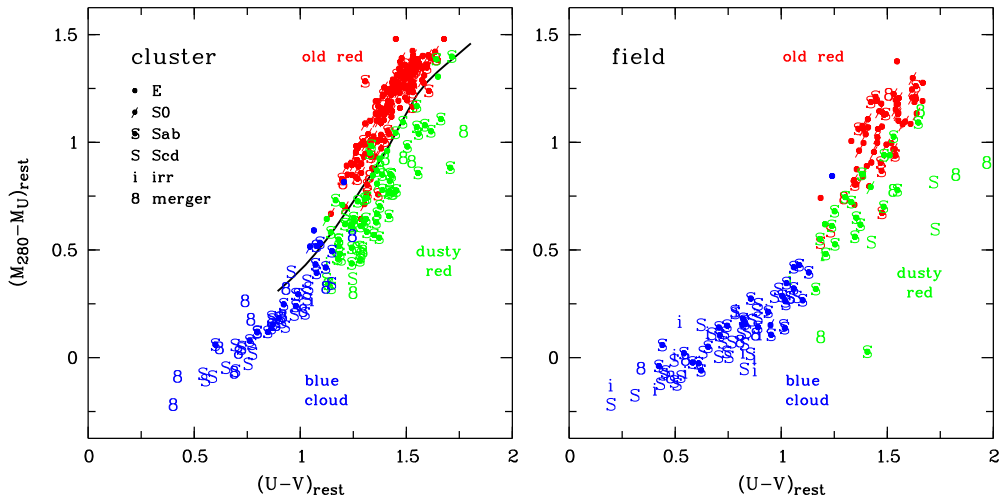}
\caption{Rest-frame colours of cluster members (left) and field galaxies (right) with $\log M_*/M_\odot > 10$: symbols indicate morphology and colours resemble SED types. The cluster contains more dusty red than blue galaxies; they are mostly early (Sa/Sb) spirals. At fixed $U$-$V$ dusty red galaxies are bluer in $M_{280}$-$M_U$ than old red ones due to young stars. A divider between old red and dusty red is shown as an age sequence with a $E_{B-V}=0.1$ (line). \label{rfcols}}
\end{figure*}

\section{Our recent work}

We split the galaxies in this sample into three categories, going beyond the classical split of red and blue galaxies: \citet{Wolf_WGM05} already described the properties of red galaxies after splitting them into old vs.~dusty. They found that dusty red galaxies in A901/2 have similar overall dust extinction levels as normal blue star-forming galaxies, while old red galaxies were defined as effectively dust-free. In a colour space where reddening due to age and dust can be distinguished, old red galaxies form a separate structure of their own, indeed a red ÔsequenceÕ, while dusty red galaxies are just a continuous tail extension of the blue cloud, and supposedly only differentiated from it by gradually lower specific star formation with redder colour.

Now, the HST data show as expected, that old red galaxies are almost exclusively E and S0 galaxies, while the blue cloud contains mostly normal spirals. The dusty red galaxies are indeed spirals with smooth arms, thus lacking already the morphological signs of star formation in form of clumps. However, they have both bluer UV slopes compared to old red galaxies and significant $24\mu$m fluxes, suggesting star formation to be present. The ratio between UV and FIR measures of star formation is correlated with the global extinction estimate from the optical SED with a slope expected from a dust law. At fixed mass these red spirals have on average 4$\times$ lower SFR than blue galaxies, though the distribution covers the whole range to full SFR suppression. We thus suggest that these objects could be spirals undergoing slow SFR quenching. 

We note that \citet{Wolf_KK04} observed spirals in the Virgo cluster that cover a similar range in SFR reduction; the dominant mode of transformation there seems to be an outside-in truncation of the star-forming disk inside these galaxies. This finding is made possible by the superior spatial resolution on galaxies as near as the Virgo cluster. \citet{Wolf_Ke08} have corroborated these findings with $24\mu$m data, showing that star formation recedes towards the centres of Virgo spirals where it is on average more obscured as well. We speculate that these galaxies may have similar SEDs and morphology to ours if observed at $z\sim 0.17$.

\begin{figure*}
\centering
\includegraphics[clip,angle=270,width=\hsize]{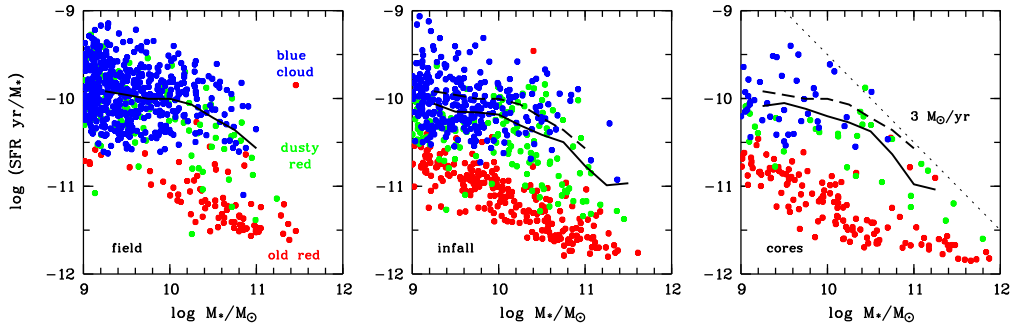}
\caption{
Star formation-mass diagram of three environments split by SED type. Lines show mean values of star-forming galaxies in mass bins (field dashed for comparison). Dusty red galaxies are most common in infall regions; at fixed mass their SFR is $\sim 4\times$ lower than that in blue galaxies.
\label{ssfrM}}
\end{figure*}

The SED-based typing in COMBO-17 already showed that dusty red galaxies were found mostly in the cluster outskirts. From there, their fraction declines both towards the cluster cores and towards the regular-density galaxy field. The present work from the STAGES data set includes typing by morphology which shows a similar picture. Here, we also resolve the trends by galaxy stellar mass, and find that red spirals are predominantly found at $\log M_*=[10,11]$. The red fraction among spirals increases both with galaxy mass and galaxy density in the environment. In particular, the red spirals in the infall regions of clusters and groups are by no means edge-on populations in contrast to field samples: in the field, 10\% of spirals are red, which is the chance fraction for a sufficiently reddening edge-on geometry; in the cluster here, however, red spirals are more common than blue spirals and are thus a genuinely physical phenomenon.

The STAGES findings are published in more detail in \citet{Wolf_W09} and match quantitatively results on red spirals found by the Galaxy Zoo project \citep{Wolf_Bam09}. While Galaxy Zoo lacks multi-wavelength data, it covers the huge volume of the entire SDSS survey and confirms that red spirals are a general cosmological phenomenon and not a localised freak event. The mass dependence of their fraction suggests a time scale for the red spiral phase that becomes longer with increased galaxy mass. Whatever process causes it, will act more slowly on larger galaxies if infall velocity did not depend on galaxy mass.

\section{Placed into context}

Altogether, we see a sequence of galaxy properties with density of their environment or with proximity to cluster centres. Given cosmological structure growth it is likely to be an evolutionary sequence even though a single-redshift cluster snapshot can not clarify the influence of redshift evolution in the progenitors of the galaxies seen: the property sequence along density intertwines secular galaxy evolution with effects due to the environment. From lower to higher density we see how the dominant galaxy type changes from blue spiral over red spiral to S0 galaxy. Along this sequence, we see the clumpiness of stellar disks as well as star formation rates decline, and UV slopes getting redder. The $U$-$V$ colours only get redder from blue to red spirals, but then change little when ageing stellar populations and clearing of dust have counter-acting effects. The mean reddening of the overall stellar population changes little from blue to red spirals, but disappears towards the S0s. The mean reddening of star formation may increase a little towards the red spiral phase after which star formation disappears.

These gradual changes observed by STAGES map out the transformation process of spirals falling into clusters more clearly now. Previously, the star formation in red spirals was not acknowledged properly. As a result, transition objects had not been identified clearly and only the disappearance of star-forming galaxies was diagnosed.

Now, Galaxy Zoo, STAGES and work on Virgo galaxies all together support the view that a large fraction of galaxies is transformed by their environment, even in the low-redshift Universe. They lose star formation slowly upon cluster infall; more quickly than due to secular evolution, but more slowly than ram-pressure stripping by hot and dense ICM would do \citep{Wolf_Mo07}. Galaxy Zoo has shown this process to be universal, STAGES has shown the remaining star formation level to be significant and stretching the whole range from little to high suppression, and the Virgo galaxies provide spatial resolution to study the processes in detail, at least for one cluster.

\section{Outlook and criticism}

However, whether the outside-in truncation of the star-forming disk is the main mode of transformation, what role it plays relative to mergers and what further modes are possible, is still unclear. We expect that a galaxy's response to its environment will depend not only on properties of the galaxy but also on those of the cluster. Also, we have not yet been able to verify which role AGN may play in this framework. We know that strong AGN are rare in our cluster, and the strong red spiral trends with density could only suggest a strong role of AGN if AGN were mainly caused by environmental aspects other than mergers. 

Next steps could involve a test whether models of disk fading are able to reproduce the evolutionary sequence and what level of morphological rearrangement is required to make S0 galaxies. Here, it is very important to look at a mass-resolved sample and keep progenitor bias and downsizing in mind: the S0s of tomorrow as derived from today's red spirals do not need to look the same as today's S0s that were made from past red spirals.

The STAGES data set presented here is available to the public, including SEDs, redshifts, stellar mass and SFR estimates in catalogue form and reduced images from COMBO-17, HST and Spitzer \citep{Wolf_G09}. We look forward to more work towards understanding environmental galaxy nurturing.

\end{document}